\documentclass[12pt]{article}
\usepackage{graphics}

\begin{document}
\baselineskip 18pt

\begin{center}
{\large\bf Testing Cosmological Models with Negative Pressure}
 
\vskip 0.5cm
 
{\large Alberto Cappi$^{1,2}$}
 
\end{center}
\medskip
\par\noindent
{\small $^1$Osservatorio Astronomico di Bologna,
via Ranzani 1, I-40127, Bologna, Italy \\ 
{\small $^2$Observatoire de la C\^ote d'Azur,
B4229,Le Mont-Gros, 06304 Nice Cedex 4, France} \\

\centerline{E--Mail: cappi@astbo3.bo.astro.it}
\vfill
\centerline{\em Astrophysical Letters and Communications, in press}

\newpage

\begin{abstract}
There is now strong evidence that the main contribution to the
cosmic energy density is not due to matter, but to another component 
with negative pressure. Its nature is still unknown: it could be
the vacuum energy, manifesting itself as a positive cosmological 
constant with $w \equiv P/\rho c^2 = -1$ ($\Lambda$CDM model), 
a spatially inhomogeneous and dynamically evolving form of energy
with $-1 < w < 0$ (quintessence, QCDM model), or a dark energy 
component with $w < -1$, such as a quantized free scalar field (VCDM model).
After presenting simple redshift--distance formulae, which are useful for 
comparing observations with theoretical predictions without numerical 
integration, I discuss the behaviour of different $w < 0$ models 
in the context of the Alcock--Paczy\'nski geometric test and the 
statistics of gravitational lensing.
\end{abstract}


\newpage

%
%
\section{Introduction}

The observations of distant Supernovae (Riess et al. 1998, 
Perlmutter et al. 1999, Riess et al. 2001) indicate that the universe is 
accelerating, as expected for a positive cosmological constant;
in this case it is possible
to reconcile a relatively high Hubble constant with an old
universe, and a low luminous+dark matter density with the flat geometry
implied by the recent CMB observations (de Bernardis et al. 2000). 
The vacuum energy density can act as a cosmological constant, 
but its low value and its magnitude comparable to the present matter density
represent an unsolved problem (see e.g. Sahni \& Starobinsky 2000, 
Carroll 2001, and references therein).
An intriguing possibility is that the cosmic energy density
is dominated by something acting as a variable cosmological ``constant"
(Ratra \& Peebles 1988); this additional component to the cosmic energy 
density, now known as ``quintessence'', would be slowly 
evolving with time and spatially inhomogeneous, and could be due
to a scalar field evolving in a potential coupled to matter through 
gravitation (Caldwell et al. 1998). 
Various authors (see Efstathiou 1999, Wang et al. 2000, Balbi et al. 2001) 
have shown that a flat model 
including a dominant component of quintessence (with equation of state 
$w_Q \sim -0.65$, see below) is consistent with observations, while
Amendola (2000) has put constraints on a  quintessence--dark matter coupling. 

The observational constraints are still quite uncertain and it is not 
possible to discriminate between the different model parameters 
(Podariu \& Ratra, 1999). Moreover,
Caldwell (1999) suggested that even models with $w < -1$
satisfy all the observational constraints; he named this 
hypothetical component ``phantom energy", as in his model there is a negative 
kinetic term in the Lagrangian of the scalar field (see also Schulz 
\& White 2001).
Nevertheless a $w < -1$  equation of state can also be obtained
assuming that the vacuum energy is due to a quantized free scalar field of 
low mass (VCDM model, Parker \& Raval 2001 and references therein).

Changing the equation of state of the dominant energy component of the universe
affects the distance--redshift relation and all the distance--dependent 
physical quantities. The effects become important in deep redshift surveys.
Unfortunately, in most cases it is not possible to find exact analytical 
expressions for the distance--redshift relation
(e.g. McVittie G.C. 1965). Models with a cosmological
constant are now well studied, and
distances and other useful quantities have
been obtained through numerical integration 
(Refsdal, Stabell \& de Lange 1967) or using elliptic integrals (Feige 1992).
An excellent approximation for $\Lambda > 0$ flat cosmologies has been found
by Pen (1999), while Kayser, Helbig \& Schramm (1997) describe a method to 
calculate cosmological distances also for inhomogeneous cosmological models.
Recently, Hamilton (2001) has also found formulae for the linear growth 
factor and its logarithmic derivative.

On the other hand, the alternative models have not been studied
extensively (but after the submission of this paper, the situation has 
continued to improve; 
see e.g. Giovi et al. 2000). As simple expressions for the 
distance--redshift relation are always useful to perform a fast analysis,
in this paper I present simple formulae for quintessence cosmological
 models. 

It is assumed that
a) the Universe is homogeneous on large--scales, so that 
Friedmann--Lema\^{\i}tre equations can be used --in the case of gravitational 
lensing filled beam (i.e. standard) distances will be adopted--, and 
b) the equation of state is constant.

In section 2 the basic equations are briefly described, and some
exact analytical formulae for the redshift--distance relation are explicitely
shown in the trivial cases of one component with $w<0$; it is also shown a 
formula with elliptic integrals for a realistic matter + quintessence 
cosmological model with $w = -2/3$, which satisfies all the presently 
available observational constraints.

In section 3 and 4 I discuss the differences among different
models with a dark energy component which are expected when applying
respectively the Alcock--Paczy\'nski test and the statistics of 
gravitational lensing. The main conclusions are in section 5.

%
%
\section{Redshift--distance relations for models with $w<0$}

\subsection{Basic equations}

For sake of clarity, the basic equations for calculating distances in 
relativistic cosmology are briefly resumed (e.g. Peebles 1993; Coles \&
Lucchin 1996, Peacock 1999; see also Hogg 1999).
Assuming that the universe is homogeneous and isotropic universe, we have the
Robertson--Walker metric:

\begin{equation}
ds^2 = c^2 dt^2 - R^2(t) \left[ dr^2 + S_k(r) 
(d\theta ^2 + \sin^2 \theta d\phi^2) \right]
\end{equation}

where $S_k(r)=r$ for $k=0$, $S_k(r)=\sin(r)$ for $k=1$ and 
$S_k(r)=\sinh(r)$ for $k=-1$,
and from Einstein's field equations the Friedmann equations are obtained:

\begin{equation}
\label{eq:friedmann1}
\left( \frac{\dot{R}}{R} \right)^2 = H^2 = 
\frac{8 \pi G}{3} \sum_i \rho_i - \frac{k c^2}{R^2}
\end{equation}

\begin{equation}
\label{eq:friedmann2}
\frac{\ddot{R}}{R} = -\frac{4 \pi G}{3} \sum_i \rho_i (1 + 3 w_i)
\end{equation}

where $\rho_i$ is the energy density of component $i$
and $w_i = P_i / \rho_i c^2$ is the corresponding equation of state.

Defining the ratios $\Omega_i \equiv \rho_i / \rho_c$, where
$\rho_c \equiv {3 H^2}/{8 \pi G}$ and
$\Omega_k \equiv - {k c^2}/(H R)^2$, we can rewrite equation 
(\ref{eq:friedmann1}) as:
$ \Omega_{tot} = \sum_i \Omega_i = 1 - \Omega_k $.

Moreover, dividing equation (\ref{eq:friedmann2}) by
equation (\ref{eq:friedmann1}) we obtain the general expression for
Sandage's deceleration parameter $q = - \ddot{R}R/\dot{R^2} =
\frac{1}{2} (1 - \Omega_k) + \frac{3}{2} \sum_i \Omega_i w_i$.

The comoving coordinate $r$ can be written as a function
of $\Omega_i$:

\begin{equation}
 \label{eq:integral}
 r = \sqrt{|\Omega_k|} \int_0 ^z \frac{dz^\prime}{E(z^\prime)} 
\end{equation}

where:

\begin{equation}
\label{eq:edz}
E(z) = \sqrt{\Omega_k (1+z)^2 +  \sum_i \Omega_i (1+z)^{3(1+w_i)} }
\end{equation}

In what follows I will consider only models with one $w < 0$ component and 
matter ($w=0$), while radiation will be neglected. 
In the following, whenever useful, a subscript will indicate the range of 
values assumed by the equation of state: $w_Q$ for values in the
range $-1 < w < 0$, and $w_V$ for values less than $-1$.

The comoving distance\footnote{$d_c(z)$ corresponds to the
transverse comoving distance $D_M$ as defined by Hogg (1999).} 
is therefore given by:

\begin{equation}
d_c(z) = R_0 S_k(r) = \frac{c}{H_0} \frac{1}{|\Omega_k|^{1/2}} S_k(r)
\end{equation}

where, as usual, the subscript $0$ indicates
the value at the present epoch.
The luminosity distance $d_L$ is simply given by $d_L = d_c (1+z)$, while
the angular distance is $d_A (z) = d_c / (1+z)$.

The lookback time is also given by an analogous integral: 

\begin{equation}
\label{eq:lookback}
t(z) = \frac{1}{H_0} \int_0 ^z \frac{1}{(1+z^\prime) E(z^\prime)}
\end{equation}

Numerical integration (see Press et al. 1992) has been used to check the 
results obtained with analytical formulae and more generally when
no analytical solution was available.

\subsection{One--component models}

Simple analytical expressions can be found only in special cases.
Calculations are clearly simplified assuming $w$ constant
(a reasonable approximation as far as the equation of state 
changes slowly with time).

For a flat universe ($\Omega_k = 0$) with only one component,
analytical solutions to the integral in equation
(\ref{eq:integral}) are very simple. Defining 
$\alpha \equiv (1+3w)/2$, we obtain:

\begin{equation}
d_c(z) = 
\left\{ \begin{array}{lcr}
\frac{c}{H_0} \frac{1}{\alpha} \left[ 1 - (1+z)^{-\alpha} \right]
                       & \rm for & w \neq -1/3 \\
\frac{c}{H_0} \ln(1+z) & \rm for & w = -1/3
\end{array}\right.
\end{equation}

Other simple solutions can be found by fixing the value of $w$.
The values $w=-1/3$ and $w=-2/3$ are of special interest because they
correspond respectively to a frustrated network of cosmic strings and of
domain walls (Bucher and Spergel 1999); morevoer, the value $w=-2/3$
is consistent with present observations.

For $w=-1/3$ we have:

\begin{equation}
\label{eq:d13}
d_c(z) = 
\left\{ \begin{array}{lcr}

\frac{c}{H_0} \frac{\sinh(\sqrt{1-\Omega_Q} \ln(1+z))}
                         {\sqrt{1-\Omega_Q}} & \rm for & 0 < \Omega_Q < 1 \\
\frac{c}{H_0} \ln (1+z) & \rm for & \Omega_Q = 1 \\
\end{array}\right.
\end{equation}

and for $w=-2/3$:

\begin{equation}
\label{eq:d23}
d_c(z) = 
\left\{ \begin{array}{lcr}
\frac{c}{H_0} \frac{1}{\sqrt{1-\Omega_Q}} \ln 
\frac{z(1-\Omega_Q)+1-\Omega_Q/2+\sqrt{[z(1-\Omega_Q)+1](1+z)(1-\Omega_Q)} }
{\sqrt{1-\Omega_Q}+1-\Omega_Q/2} & 0 < \Omega_Q < 1 \\
2 \frac{c}{H_0} (\sqrt{1+z} -1) & \Omega_Q = 1 \\
\end{array}\right.
\end{equation}

Figure 1a shows the distance--redshift relation for flat cosmological models
with one component having different equations of state, 
and assuming $H_0 = 65$ km/s Mpc$^{-1}$. It is clear that a more negative
equation of state implies a larger distance at a given redshift, and that the
difference increases with redshift.

%
%
\begin{figure}
\resizebox{\hsize}{!}{\includegraphics{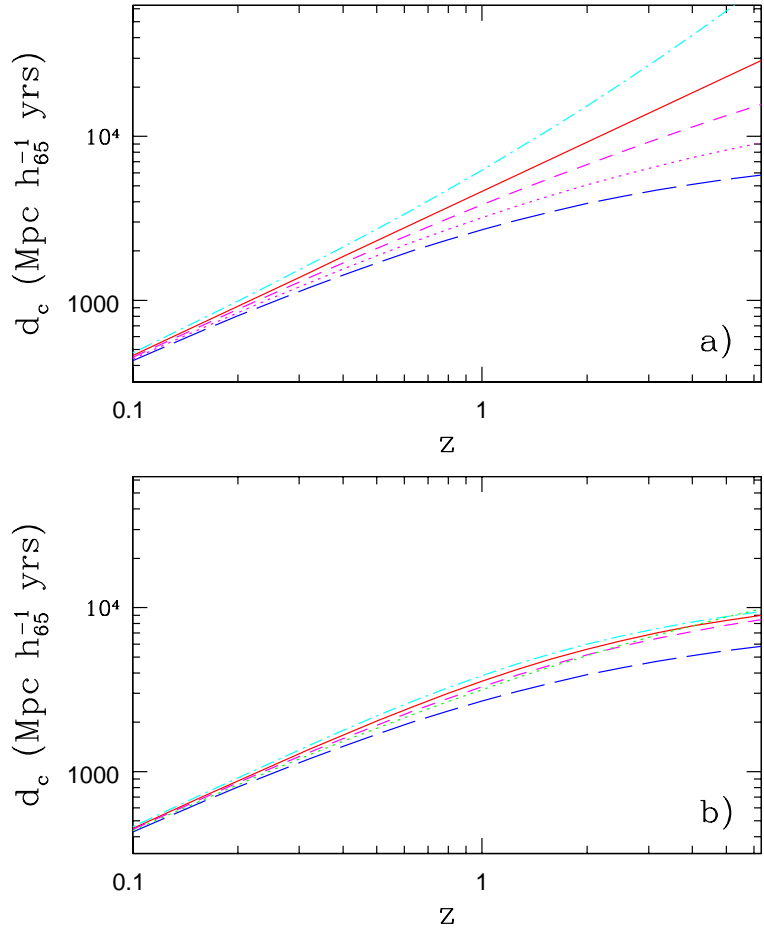}}
\caption{a) Comoving distances for flat cosmological models with one 
component. Dotted dashed line: $w=-3/2$;
solid curve: $w=-1$; dashed curve: $w=-2/3$; dotted curve: $w=-1/3$;
long dashed curve: Einstein--de Sitter model.
b) Comoving distances for an $\Omega_M = 0.3$,
open model (dotted curve), the Einstein--de Sitter model (long--dashed curve),
and 3 flat models, with $\Omega_M = 0.3$ and
an additional negative pressure component $\Omega_X=0.7$ with the following
equation of state:
dashed--dotted curve: $w=-3/2$;
solid curve: $w=-1$;
dashed curve: $w=-2/3$.
}
\label{fig1}
\end{figure}

\subsection{A formula for the ``standard" quintessence model} 

Numerical integration of equation (\ref{eq:integral}) is necessary
in the general --and realistic-- case of a two--component model 
including both matter and dark energy.
In the case of {\em flat} cosmological models,
the comoving distance of a galaxy at a redshift $z$ 
can be written using the hypergeometric function $_2F_1(a,b;c;z)$:

\begin{eqnarray}
d_c(z) =  2 \frac{c}{H_0} \frac{1}{\sqrt{1-\Omega_X}}
\{ \ _2F_1 \left[ -\frac{1}{6w}, \frac{1}{2}, 1-\frac{1}{6w}, 
- \frac{\Omega_X}{1-\Omega_X} \right] \nonumber \\
- \frac{1}{\sqrt{1 + z}} 
\ _2F_1 \left[ -\frac{1}{6w}, \frac{1}{2}, 1-\frac{1}{6w}, 
- \frac{\Omega_X}{1-\Omega_X} (1 + z)^{3w} \right] \}
\end{eqnarray}

where $\Omega_X$ refers to the dark energy contribution.
The above relation is consistent with the formula found by Torres and Waga 
(1996) for angular diameter distances.
Assuming $w=-2/3$ it is possible to further simplify the above expression,
using only the incomplete elliptic integral of the first kind.
A simpler expression can be of practical utility as 
this equation of state is consistent with the restricted range permitted 
by present observations 
(the best parameters for a flat universe are $\Omega_M=0.33 \pm 
0.05$ and $w_Q = 0.65 \pm 0.07$ according to Wang et al. 1999).
 
The formula is the following:

\begin{equation}
d_c(z) =
\left\{ \begin{array}{llll}
C
\left[ u \left(g(0),\frac{1}{\sqrt{2}} \right) - 
u \left(g(z),\frac{1}{\sqrt{2}} \right) \right]
& 0 < \Omega_Q \le 0.5 & & \\
C \left[ u \left(g(z),\frac{1}{\sqrt{2}} \right) - 
u \left(g(0),\frac{1}{\sqrt{2}} \right) \right]
& 0.5 < \Omega_Q < 1, & z \le \sqrt{\frac{\Omega_Q}{1-\Omega_Q}}-1 \\
C \left[ 2 u \left(1,\frac{1}{\sqrt{2}} \right) - 
u \left(g(0),\frac{1}{\sqrt{2}} \right)-
u \left(g(z),\frac{1}{\sqrt{2}} \right) \right]
& 0.5 < \Omega_Q < 1, & z > \sqrt{\frac{\Omega_Q}{1-\Omega_Q}}-1 \\
\end{array} \right.
\end{equation}

where:

\begin{equation}
C = \frac{c}{H_0}\frac{1}{(\Omega_Q(1-\Omega_Q))^{1/4}}
\end{equation}

\begin{equation}
g(z) = 2 \frac{ \left(\frac{\Omega_Q}{1-\Omega_Q} \right)^{1/4} \sqrt{1+z} } 
{ \sqrt{ \frac{\Omega_Q}{1-\Omega_Q} } +1 + z }
\end{equation}

and 

\begin{equation}
u(g, \frac{1}{\sqrt{2}}) = 
\int_0 ^g \frac{1}{ \sqrt{(1-t^2)(1-t^2/2)} } dt
\end{equation}

is the incomplete elliptic integral of the first kind, which can
be easily computed with one of the standard routines available 
in many mathematical 
packages\footnote{For example {\em EllipticF} in Maple V
or {\em ellf} in {\em Numerical Recipes}: notice that {\em ellf}
requires the quantity asin($g$) instead of $g$.}.
$u(1,1/ \sqrt{2}) = K(1/ \sqrt{2})$ corresponds to the 
complete elliptic integral of the first kind, and
$u(g(0),1/ \sqrt{2})$ depends only on $\Omega_Q$: when $\Omega_Q =0.5$,
$u(g(0),1/ \sqrt{2}) = K(1/ \sqrt{2})$.

Figure 1b shows $d_c(z)$ for the Einstein--de Sitter model,
an open model with $\Omega_M=0.3$, 
and other 3 flat models with $\Omega_M = 0.3$
and a negative pressure component with $w=-2/3$ (QCDM), $w=-1$
($\Lambda$CDM), and $w = -3/2$ (VCDM).
A synthesis of their main properties is given in table (1).
A comparison between figures 1a and 1b shows clearly that
the difference between models becomes small in the presence of 
matter. For example, up to $z \sim 3$ the curve of 
the open model is very similar to that of the quintessence model with 
$w=-2/3$. 

%
%
\begin{table}
\label{tab:models}
\caption[]{Properties of different cosmological models}
\begin{flushleft}
\begin{tabular}{lrrrrr}
\hline
  & (1,0,0) & (0.3,0,0) & (0.3,0.7,-2/3) & (0.3,0.7,-1) & (0.3,0.7,-3/2) \\
\hline
$q_0$                             & 0.5  & 0.15 & -0.2 & -0.55 & -1.07 \\
Age ($10^9$h$_{65}^{-1}$ yrs)     & 10.3 & 12.2 & 13.6 & 14.5  & 15.4  \\ 
$z(\theta_{min})$                 & 1.25 & 1.90 & 1.59 & 1.61  & 1.55  \\
$z(F_{max})$                      & --   & 1.02 & 3.29 & 1.48  & 0.79  \\     
$A(F_{max})$                      & 1.00 & 1.06 & 1.24 & 1.23  & 1.22  \\
\hline
\end{tabular}
\end{flushleft}
\end{table}

%
%
\begin{figure}
\resizebox{\hsize}{!}{\includegraphics{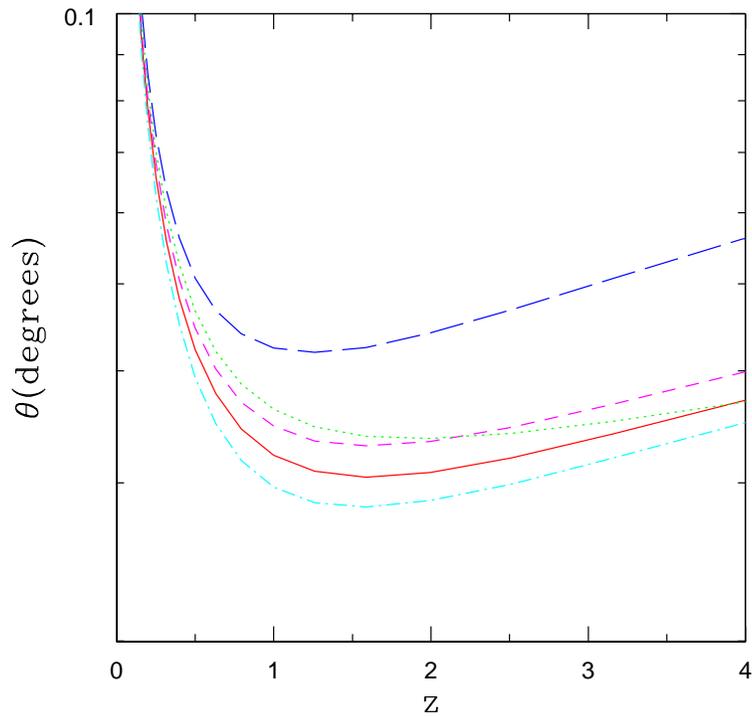}}
\caption{a) Angular diameter corresponding to 1 Mpc 
(for $H_0=65$ km/s/Mpc) as a function of redshift for the same models
as in figure 1b. Dotted curve: $\Omega_M = 0.3$,
open model; long--dashed curve: Einstein--de Sitter model;
3 flat models, with $\Omega_M = 0.3$ and
an additional negative pressure component $\Omega_X=0.7$ with the following
equation of state:
dashed--dotted curve: $w=-3/2$; 
solid curve: $w=-1$;
dashed curve: $w=-2/3$.}
\label{fig2}
\end{figure}

Another interesting quantity is the angular 
diameter.
The angular diameter of an object with physical diameter $D$ is
$ \theta = \frac{D (1+z)}{d_c(z)} = \frac{D (1+z)^2}{d_L(z)} $.
Figure 2 shows the angular diameters corresponding to 
1 $h_{65} ^{-1}$ Mpc for the same 4 models as in figure 1b. Also in this case 
the differences between models with negative pressure are not very large,
and it is apparent the similarity between the open and standard quintessence 
models.
A detailed discussion of the angular size in the context of quintessence 
models can be found in Lima \& Alcaniz (2000).

%
%
\section{The Alcock--Paczy\'nski test}
 
Alcock \& Paczy\'nski (1979) showed that, at least in principle, it is possible
to detect the presence of the cosmological constant $\Lambda$ in a redshift 
survey through a geometric test which is independent of galaxy evolution. 
This method is based on the
fact that the spatial distribution of galaxies obtained from the 
distance--redshift relations valid for an
Einstein--de Sitter model is significantly distorted if the universe 
has a $\Lambda$ component.
Therefore a measure of the apparent anisotropy in known isotropic 
structures would give us the value of the cosmological constant.
Phillips (1994) applied this method to the orientation of quasar pairs, without
conclusive results. His work has been extended to the quasar correlation 
function by Popowski et al. (1998).
The main problem is that redshift distortions are also induced
by peculiar velocities --$\beta$--distortion 
($\beta \equiv \Omega_M ^{0.6}/b$) on linear scales and fingers of God on 
small scales--, but the method might be successfully applied to future
redshift surveys, with the simultaneous extraction of $\Lambda$
and $\beta$ from anisotropic power--spectrum data, 
as discussed by Ballinger, Peacock \& Heavens (1996).

%
%
\begin{figure}
\resizebox{\hsize}{!}{\includegraphics{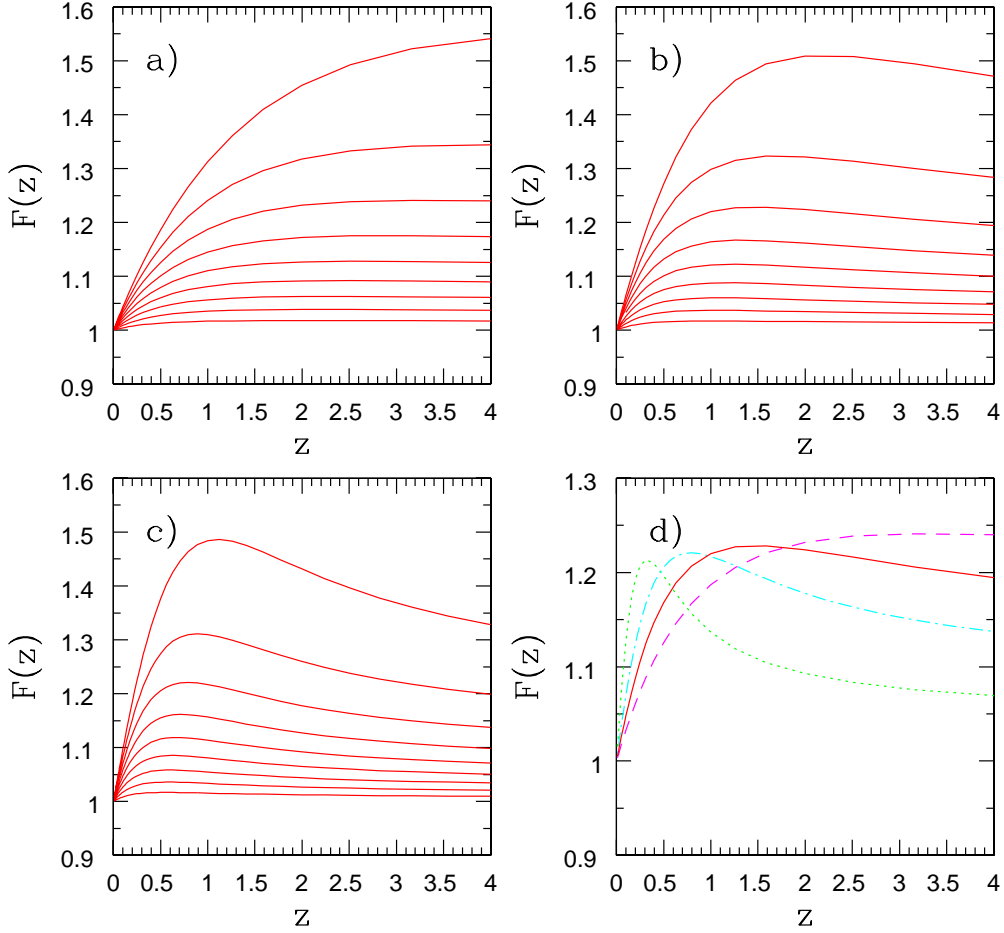}}
\caption{Flattening factor $F(z)$ for different flat models; in figures a--c,
the value of $\Omega_M$ varies from $0.1$ (top curve) to 
$\Omega_M=0.9$ (bottom curve) in steps of 0.1.
a) QCDM ($\Omega_M+\Omega_Q=1$, $w_Q=-2/3$);
b) $\Lambda$CDM ($\Omega_M+\Omega_\Lambda=1$);
c) VCDM ($\Omega_M+\Omega_V=1$, $w_V=-3/2$);
d) 4 flat models with $\Omega_M=0.3$. 
Dashed curve: QCDM model with $w_Q=-2/3$;
solid curve: $\Lambda$CDM  model; 
dashed--dotted curve: VCDM model with $w_V=-3/2$;
dotted curve: VCDM model with $w_V=-3$.
}
\label{fig3}
\end{figure}

The distortion can be quantified by a flattening factor $F(z)$
 relative to the EdS model. In our notation, it is defined as:

\begin{equation}
F(z) \equiv 2 \frac{c}{H_0} \frac{(1+z)^{3/2}-(1+z)}{d_c(z)E(z)}
\end{equation}
$F > 1$ indicates a flattening
along the line of sight (see Ballinger et al. 1996).
Only in special cases $F(z)$ can be expressed analytically:
for example, for a flat quintessence--only model ($\Omega_Q=1$) and 
$w_Q=-2/3$, we have $F(z)=\sqrt{1+z}$, to be compared to the $\Omega_\Lambda=1$
case, with $F(z)=2(1+z)[\sqrt{(1+z)}-1]/z$.

Note that the formula valid at first order in $z$ given by
Ballinger et al. (1996) can be generalized to quintessence models:

\begin{equation}
F(z) = 1+\frac{1}{4}[1-\Omega_M-(1+3w_Q)\Omega_Q] z + O(z^2)
\end{equation}

Therefore, when $w_Q=-2/3$, the test is a direct measure of
the difference $\mid \Omega_M-\Omega_Q \mid$.

In figure \ref{fig3}a--c it is shown $F(z)$ for flat models with
$w=-2/3$, $w=-1$, and $w=-3/2$, while
figure \ref{fig3}d shows 4 flat models with $\Omega_M = 0.3$.

There are three main systematic effects due to the different equation of 
state which appear clearly in figure \ref{fig3}d. 

With a more negative $w$ we have:

\begin{itemize}
\item{a)} the amplitude of the maximum anisotropy becomes (slightly) smaller;
\item{b)} the maximum of $F(z)$ shifts towards smaller redshifts;
\item{c)} after the maximum, $F(z)$ decays faster.
\end{itemize}

With a realistic fraction of matter the amplitude of the geometric distortion 
is not very large, as it amounts to about 20\%.
After the first analysis of Boomerang results,
McGaugh (2000) stressed that a universe with 
$\Omega_\Lambda \sim 1$ and $\Omega_M = \Omega_b$ would be
consistent with results from the observations of distant type Ia supernovae
and with the small amplitude of the second peak in the power spectrum of 
anisotropies in the CMB, if instead of CDM one adopted MOND (Milgrom 1983). 
In this case, we would expect a larger geometrical distortion,
about 30\% at $z=0.5$ and 50\% at $z = 1$. However,
the more extended analysis of Boomerang results, which has recently 
identified the first three peaks in the CMB angular power spectrum, 
shows that a standard flat model with $\Omega_\Lambda=0.7$ and
$\Omega_M=0.3$ is consistent with primordial nucleosynthesis 
(Netterfield et al. 2001).

An interesting aspect of the Alcock--Paczy\'nski test applied to
the VCDM models is that the maximum distortion is shifted towards
relatively low redshifts: $z_{max} \sim 0.32$ if $w_V = -3$, 
and even $z_{max} \sim 0.08$ in an extreme case $w_V = -10$.
This well--defined maximum at moderate redshifts could be detected in 
on--going galaxy redshift surveys, or at least it should be possible to fix a 
lower limit to the equation of state, depending on our ability to disentangle
the geometric distortion from the redshift distortion due to peculiar 
velocities.

%
%
\section{Statistics of gravitational lensing}

Gravitational lensing represents an important test of cosmological models, 
as it is sensitive to the presence of a cosmological constant 
(Fukugita et al. 1992) or to an equivalent component with negative pressure.
Here we simply aim to examine the relative trend for the different models,
without going in much detail (among the more recent studies see
e.g. Kochanek 1996, Chiba \& Yoshii 1999).

The differential probability $d\tau$ of a line of sight
intersecting a lensing galaxy (modelled as a Singular
Isothermal Sphere) at redshift $z_L$ in the redshift interval 
$dz_L$ is:

\begin{equation}
d\tau = f \frac{(1+z_L)^3}{E(z_L)} \left(\frac{c}{H_0}\right)^{-2} 
\frac{d_A(0,z_L) d_A(z_L,z_S)}{d_A(0,z_S)} dz_L
\end{equation}

For flat models, we have the analytical relation:

\begin{equation}
\tau(z_S) = \frac{f}{30} d_c(z_S)^3 \left(\frac{c}{H_0} \right)^{-3}
\end{equation}

where $f$ measures the effectiveness of the lens in producing double images
(Turner, Ostriker \& Gott 1984); here its value is estimated assuming a
Schechter luminosity function:

\begin{equation}
f = 
\frac{16 \pi ^3}{c H_0 ^3} 
\phi^* {\sigma ^{*}} ^4 \Gamma \left(\alpha+\frac{4}{\gamma}+1 \right)
\end{equation}

Figure \ref{fig4} shows the optical depth as a function of the redshift of 
the source, calculated using the angular diameter distance.
With the above relations, it is possible to compare the predictions of 
gravitational lensing for various cosmological models to observations.
Torres \& Waga (1996) have examined the behaviour of a model with
$w=-1/3$. Here their analysis is extended to include the more popular value
$w=-2/3$ for quintessence, and the case $w=-3/2$ for VCDM.
I have used the same catalog, the HST Snapshot Survey (see Maoz et al. 1993), 
which has detected a number between 3 and 6 of gravitational lensing candidates
out of 502 observed quasars. Magnification bias has been taken into account 
as in Maoz \& Rix (1993), and I have adopted $\phi^*=0.015$ h$^{3}$Mpc$^{-3}$,
 $\alpha = -1.2$, $\sigma^* = 200$ km/s and $\gamma=4$.

%
%
\begin{figure}
\resizebox{\hsize}{!}{\includegraphics{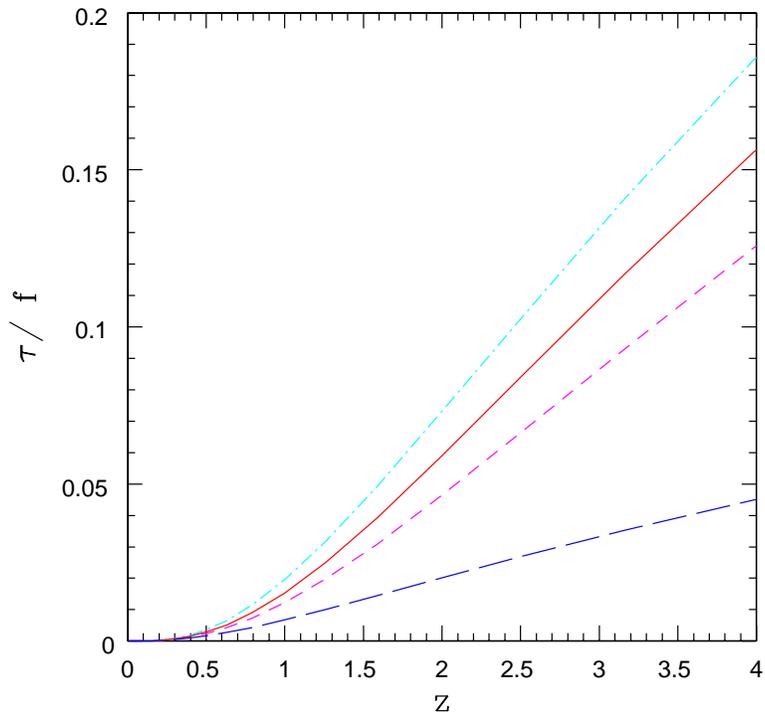}}
\caption{Optical depth for different flat models with
$\Omega_M=0.3$. From top to bottom:
dashed--dotted curve: VCDM model with $w_V=-3/2$;
solid curve: $\Lambda$ model ($w_\Lambda=-1$);
dashed curve: quintessence model with $w_Q=-2/3$;
long--dashed curve: Einstein--de Sitter model.
}
\label{fig4}
\end{figure}

Table 2 resumes the main results, assuming 4 genuine lensing events in the 
HSTSS. It is well known that $\Lambda > 0$ models predict a too large number 
of lensed quasars, as the optical length increases with
$\mid w \mid$ (as shown in figure 4; see also Zhu 2000). 
While the $w_Q=-1/3$ quintessence model
considered by Torres \& Waga (1996) could still be consistent with
the data, the same is not true for the $w_Q=-2/3$ model, and VCDM models 
perform still worse than the $\Lambda$ model.
For VCDM models, more acceptable results can be obtained if the relative
contribution of matter is increased. 
For example, in a fine--tuned VCDM universe with equal 
contributions of matter and dark energy ($\Omega_M=0.5$ and $\Omega_V=0.5$)
the number of lenses becomes comparable to the quintessence $w_Q=-2/3$ model
(see table 2). A large matter density results also from 
the analysis of radio--selected gravitational lenses and optically selected 
lensed quasars, which gives $\Omega_M > 0.38$ assuming a flat universe and a
nonzero cosmological constant (Falco, Kochanek \& Mu\~noz 1998).

New surveys should clarify the reasons of this cosmic ``discordance'' 
with respect to the increasing observational evidence that 
the energy density of the universe is dominated by a component with negative 
pressure: it could be due to systematic errors and uncertainties in the 
modeling of gravitational lenses (Cheng \& Krauss 2000).

%
%
\begin{table}
\caption[]{Predicted number of lensed quasars and corresponding probabilities.}
\begin{flushleft}
\begin{tabular}{lrrrrr}
\hline
& (1,0,0) & (0.3,0.7,-1) &  (0.3,0.7,-2/3) & (0.3,0.7,-3/2) & (0.5,0.5,-3/2) \\
\hline
Lens no.    & 3.7     &    10.8 &    8.6  &  13.5     & 8.0 \\
Prob.       & 0.193   & $0.012$ & $0.041$ & $0.002$ & $0.058$\\
\hline
\end{tabular}
\end{flushleft}
\end{table}

%
%
\section{Conclusions}

In this paper various aspects of negative--pressure cosmological
models have been analysed:

a) simple redshift--distance formulae (the equivalent of Mattig formulae)
have been derived for quintessence cosmological models;

b) it has been shown that the Alcock--Paczy\'nski test is particularly 
sensitive to $w<-1$ models, where the distortion has a maximum at a 
relatively low redshift; 
in the case of a quintessence model with $w=-2/3$, this test measures directly 
the difference $\mid \Omega_Q - \Omega_M \mid$.

c) as the more negative is $w$, the larger is the 
optical depth, a quintessence cosmological model with $w=-2/3$  predicts 
a lower number of lensed quasars --and is therefore more consistent with 
observations-- than the corresponding $\Lambda$ model, 
while the $w=-3/2$ model gives an even larger number of expected lenses, 
which can be reduced by increasing $\Omega_M$.

A more refined analysis is obviously required to test 
models with negative pressure and their consistency with the full set of 
observational data (for example, adopting 
tracking models with a variable equation of state, see Benabed \& 
Bernardeau 2001). 
 
Oncoming deep redshift surveys, such as the VLT--VIRMOS Deep Survey 
(Le F\`evre et al. 2001) and DEEP (Davis \& Newman 2001),
will sample distances larger than $z \sim 1$, so that it should be 
possible to obtain constraints to the cosmological parameters which will be 
complementary to other measurements.

%
%

\end{document}